\documentclass[twocolumn,showpacs,prb,superscriptaddress]{revtex4}

\usepackage{graphicx}
\usepackage{dcolumn}
\usepackage{bm}
\usepackage{color} 

\begin{document}

\title{
Infrared study on the electronic structure of alkaline-earth-filled skutterudites $AT_{4}$Sb$_{12}$ ($A$~=~Sr, Ba; $M$~=~Fe, Ru, Os)
}
\author{Shin-ichi Kimura}
 \altaffiliation[Electronic address: ]{kimura@ims.ac.jp}
\affiliation{
UVSOR Facility, Institute for Molecular Science, Okazaki 444-8585, Japan
}
\affiliation{
School of Physical Sciences, The Graduate University for Advanced Studies (SOKENDAI), Okazaki 444-8585, Japan
}
\author{Hojun Im}
 \altaffiliation{
Present address: Department of Physics, Sungkyunkwan University, Korea.
}
\affiliation{
School of Physical Sciences, The Graduate University for Advanced Studies (SOKENDAI), Okazaki 444-8585, Japan
}
\author{Takafumi Mizuno}
\affiliation{
School of Physical Sciences, The Graduate University for Advanced Studies (SOKENDAI), Okazaki 444-8585, Japan
}
\author{Shota Narazu}
\affiliation{
Department of Quantum Matter, ADSM, Hiroshima University, Higashi-Hiroshima 739-8530, Japan
}
\author{Eiichi Matsuoka}
 \altaffiliation{
Present address: Department of Physics, Tohoku University, Sendai, Japan
}
\affiliation{
Department of Quantum Matter, ADSM, Hiroshima University, Higashi-Hiroshima 739-8530, Japan
}
\author{Toshiro Takabatake}
\affiliation{
Department of Quantum Matter, ADSM, Hiroshima University, Higashi-Hiroshima 739-8530, Japan
}
\affiliation{
Institute for Advanced Materials Research, Hiroshima University, Higashi-Hiroshima 739-8530, Japan
}

\date{\today}

\begin{abstract} 
The optical conductivity [$\sigma(\omega)$] spectra of alkaline-earth-filled skutterudites with the chemical formula $A^{2+}M_{4}$Sb$_{12}$ ($A$~=~Sr, Ba, $M$~=~Fe, Ru, Os) and a reference material La$^{3+}$Fe$_{4}$Sb$_{12}$ were obtained and compared with the corresponding band structure calculations and with calculated $\sigma(\omega)$ spectra to investigate their electronic structures.
At high temperatures, the energy of the plasma edge decreased with the increasing valence of the guest atoms $A$ in the Fe$_{4}$Sb$_{12}$ cage indicating hole-type conduction.
A narrow peak with a pseudogap of 25~meV was observed in SrFe$_{4}$Sb$_{12}$, while the corresponding peak were located at 200 and 100~meV in the Ru- and Os-counterparts, respectively.
The order of the peak energy in these compounds is consistent with the thermodynamical properties in which the Os-compound is located between the Fe- and Ru-compounds.
This indicated that the electronic structure observed in the infrared $\sigma(\omega)$ spectra directly affects the thermodynamical properties.
The band structure calculation implies the different electronic structure among these compounds originates from the different $d$ states of the $M$ ions.
\end{abstract}


\maketitle
\section{Introduction}

\begin{figure}[t]
\begin{center}
\includegraphics[width=0.4\textwidth]{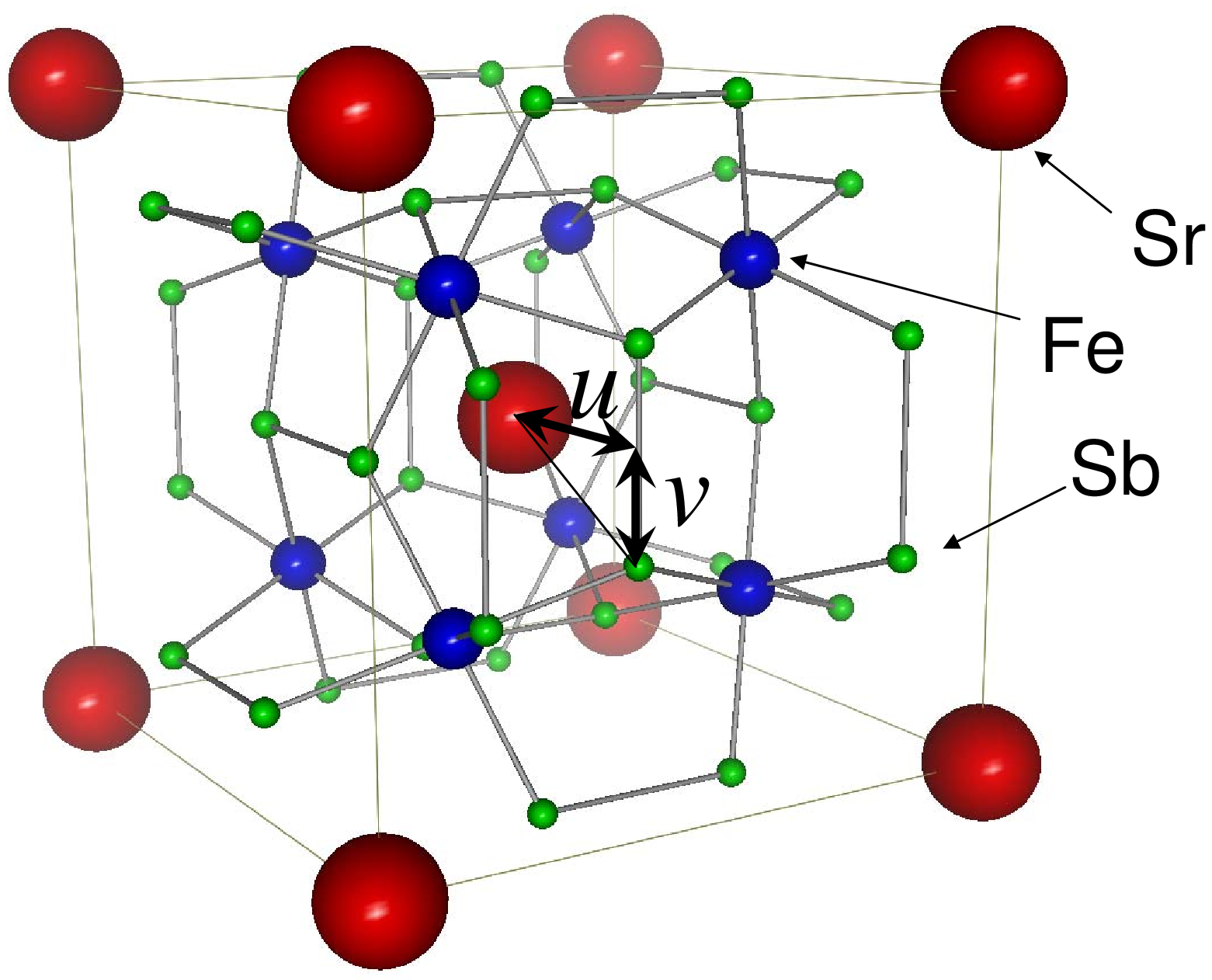}
\end{center}
\caption{
(Color online)
The crystal structure of SrFe$_{4}$Sb$_{12}$.
The unit cell is marked by lines.
The positional parameters $u$ and $v$ of an Sb atom are denoted.
}
\label{crystal}
\end{figure}

Rare-earth-filled skutterudites have recently attracted attention due to their various unique physical properties such as the magnetic field-induced quadrupole order in PrOs$_{4}$Sb$_{12}$~\cite{Aoki03}, the pressure-induced superconducting phase in PrRu$_{4}$P$_{12}$,~\cite{Miyake04} new ordered states in high magnetic fields in PrFe$_{4}$P$_{12}$,~\cite{Tayama04} a multipole ordered state in SmRu$_{4}$P$_{12}$,~\cite{Hachi06} and the magnetic field-independent heavy fermion state in SmOs$_{4}$Sb$_{12}$~\cite{Sato06} to name a few.
These unconventional physical properties indicate that the combination of the guest rare-earth ions and the other two elements of the cage produce unique physical environments due to the strong hybridization between these ions.
In the filled skutterudites case, the crystal structure shown in Figure~\ref{crystal} (body-centered-cubic structure with $Im\bar{3}$ ($T_h^5$)~\cite{THK2001}), $AM_{4}X_{12}$ ($A$: guest atom, $M$: transition metals, $X$: pnictogen), is the key for this strong hybridization.
The crystal structure makes nano-sized cages composed of $M_{4}X_{12}$.
The guest atoms $A$ are located in these cages.~\cite{HK03}
The strong hybridization between the $A$ ion and the $M_{4}X_{12}$ cage is believed to be the origin of the various unique physical properties mentioned above.
In addition, the electronic structure of the $M_{4}X_{12}$ frame is also important in determining the bulk physical properties, particularly given the magnetic moment of the $M$ ion.
For example, in the case of non-magnetic guest atoms, alkali-metal-filled iron antimony skutterudites (KFe$_{4}$Sb$_{12}$ and NaFe$_{4}$Sb$_{12}$) exhibit weak itinerant ferromagnetism due to the Fe~$3d$ electrons.~\cite{ALJ03,ALJ04,Sheet05}
These skutterudites also produce iron-based heavy quasiparticles as a result.~\cite{Joerg06,sk06-1}
The positional parameters $u$ and $v$ of the Sb atoms denoted in Figure~\ref{crystal} change depending on the different $A$ and $M$ atoms.
These positional parameters also affect the electronic structure.~\cite{Koga2005}
Given the positional dependence, it is important to ascertain the relationship between these parameters and the electronic structure as well as the physical properties.

Alkaline-earth-filled antimony skutterudites $AM_{4}$Sb$_{12}$ ($A$~=~Ca, Sr, Ba, $M$~=~Fe, Ru, Os) have been vigorously investigated.~\cite{Taka06}
Among these materials, $A$Fe$_{4}$Sb$_{12}$ materials have an electrical resistivity with a shoulder at 70~K with a quadratic dependence upon decreasing temperatures, a thermopower with a local maximum at around 50~K, an electronic specific heat coefficient $\gamma$ of 100~mJ/mol~K$^{2}$ and a maximum magnetic susceptibility at 50~K, which commonly implies the presence of low energy spin fluctuations described by the self-consistent renormalization (SCR) theory.~\cite{Matsuoka05,Matsumura05}
The ratio between the enhanced coefficient $A$ of the quadratic electrical resistivity ($\rho$ = $AT^2$) and $\gamma$ is close to the Kadowaki-Woods value [1.0 $\times$ 10$^{-5} \mu \Omega$ cm K$^{-2}$/(mJ/mol~K$^{2}$)$^{2}$].
This phenomenon originates from the presence of heavy quasiparticles, observed in an infrared reflectivity experiment, arising from the Fe $3d$ electrons.~\cite{sk06-1}
In comparison, alkaline-earth-filled ruthenium antimony skutterudites, $A^{2+}$Ru$_4$Sb$_{12}$, exhibit normal metallic physical properties with a small electrical specific heat coefficient, $\gamma$~=~10~mJ/mol~K$^{2}$.
The electrical resistivities of $A^{2+}$Os$_4$Sb$_{12}$ compounds exhibit a pronounced shoulder at 100~K, a magnetic susceptibility with a logarithmic temperature dependence and a moderately high $\gamma$ is moderately high value of 45~mJ/mol~K$^{2}$.~\cite{Matsuoka06} 
The thermodynamical properties of the Os-compounds imply that they are located between the Fe- and Ru-compounds.

To clarify the differences in the physical properties in $A^{2+}M_{4}$Sb$_{12}$ compounds ($A$~=~Sr, Ba, $M$~=~Fe, Ru, Os), the optical conductivity [$\sigma(\omega)$] spectra of these compounds were obtained and compared with the band structure calculations.
The specific purpose of this comparison is to observe and differences between the experimental and calculated $\sigma(\omega)$ spectra using the lattice constants and the positional parameters experimentally obtained by X-ray diffraction.
To check the effects of the positional parameters of the Sb atoms, the band structures of SrFe$_4$Sb$_{12}$ were also calculated using the positional parameters of the Ru-counterpart.
In the next section, the experimental and analytical methods including the theoretical calculations are explained.
In Section~3, the experimental and the calculated $\sigma(\omega)$ spectra are compared and which parameter is more influential regarding the electronic structure near the Fermi level ($E_{\rm F}$) is discussed.
The results are summarized in Section~4.

\section{Experimental and Band Calculation Methods}

High density polycrystalline $AM_{4}$Sb$_{12}$ samples were synthesized using a spark-plasma sintering technique previously reported.~\cite{Matsuoka05}
The near-normal incident optical reflectivity [$R(\omega)$] spectra were acquired from well-polished samples by using 0.3~$\mu$m grain-size Al$_{2}$O$_{3}$ wrapping film sheets.
Martin-Puplett and Michelson type rapid-scan Fourier spectrometers  (JASCO Co. Ltd., FARIS-1 and FTIR610) were used at photon energies ($\hbar \omega$) of 2.5~--~30~meV and 5~meV~--~1.5~eV, respectively, at sample temperatures in the range of 7~--~300~K using a closed cycle-helium cryostat for FTIR610 and a liquid-helium flow-type cryostat for FARIS-1.
To obtain the absolute $R(\omega)$ values, the samples were evaporated {\it in-situ} with gold, whose spectrum was then measured as a reference.
To obtain $\sigma(\omega)$ spectra via the Kramers-Kronig analysis (KKA), $R(\omega)$ was measured at 300~K over the energy range of 1.5~--~30~eV at the synchrotron radiation beam line 7B of UVSOR-II, at the Institute for Molecular Science.~\cite{BL7B}
Since $R(\omega)$ does not significantly change with temperature above 1.5~eV, the $R(\omega)$ spectra above 1.5~eV were connected to the $R(\omega)$ spectra at other temperatures in the energy range below 1.5~eV.
In the energy ranges below 2.5~meV and above 30~eV, the spectra were extrapolated using the Hagen-Rubens function [$R(\omega)~=~1-(2\omega/\pi \sigma_{DC})^{1/2}$] and the $R(\omega) \propto \omega^{-4}$ relationship, respectively.~\cite{DG02}
Here $\sigma_{DC}$ denote the direct current conductivity.
After constructing $R(\omega)$ in the energy region from zero to infinity, the KKA was performed to obtain the $\sigma(\omega)$ spectrum.

\begin{table}[t]
\caption{
The lattice constants and the positional parameters ($u, v$) of Sb atoms in $AM_{4}$Sb$_{12}$ ($A$~=~Sr, Ba, La, $M$~=~Fe, Ru, Os) used in the band calculations.
}
\label{table1}
\begin{center}
\begin{tabular}{@{\hspace{\tabcolsep}\extracolsep{\fill}}cccc} \hline
Compound & lattice constant (pm) & $u$ & $v$ \\ \hline
SrFe$_{4}$Sb$_{12}$ & 918.126 & 0.338393 & 0.160343 \\
SrRu$_{4}$Sb$_{12}$ & 928.91  & 0.343316 & 0.158496 \\
SrOs$_{4}$Sb$_{12}$ & 933.134 & 0.34154  & 0.15660  \\
BaRu$_{4}$Sb$_{12}$ & 931.51  & 0.344393 & 0.159263 \\
BaOs$_{4}$Sb$_{12}$ & 933.507 & 0.34174  & 0.157507 \\
LaFe$_{4}$Sb$_{12}$ & 913.952 & 0.336966 & 0.160425 \\ \hline
\end{tabular}
\end{center}
\medskip
\end{table}
The band structure calculation was performed using the full potential linearized augmented plane wave plus the local orbital (LAPW + lo) method including spin orbit coupling implemented in the {\sc Wien2k} code.~\cite{WIEN2k}
The no overlapping muffin-tin (MT) sphere radii of 2.50, 2.50 and 2.27 Bohrs radii were used for the Sr, Fe and Sb atoms in SrFe$_{4}$Sb$_{12}$, respectively.
The radii of all other atoms in other compounds were set to similar values.
The value of $R_{MT}K_{max}$ (the smallest MT radius multiplied by the maximum $k$ value in the expansion of plane waves in the basis set), which determines the accuracy of the basis set used, was set to 7.0.
The total number of Brillouin zones was sampled with 4000~$k$ points.
The lattice constants and positional parameters reported previously~\cite{Kaiser99} for Fe- and Ru-compounds were used.
For Os-compounds, the parameters experimentally obtained by X-ray diffraction were used.~\cite{Tsubota06}
The lattice constants and the positional parameters for an Sb atom that were used in this paper are listed in Table~\ref{table1}.

\section{Results and Discussion}

\begin{figure}[t]
\begin{center}
\includegraphics[width=0.45\textwidth]{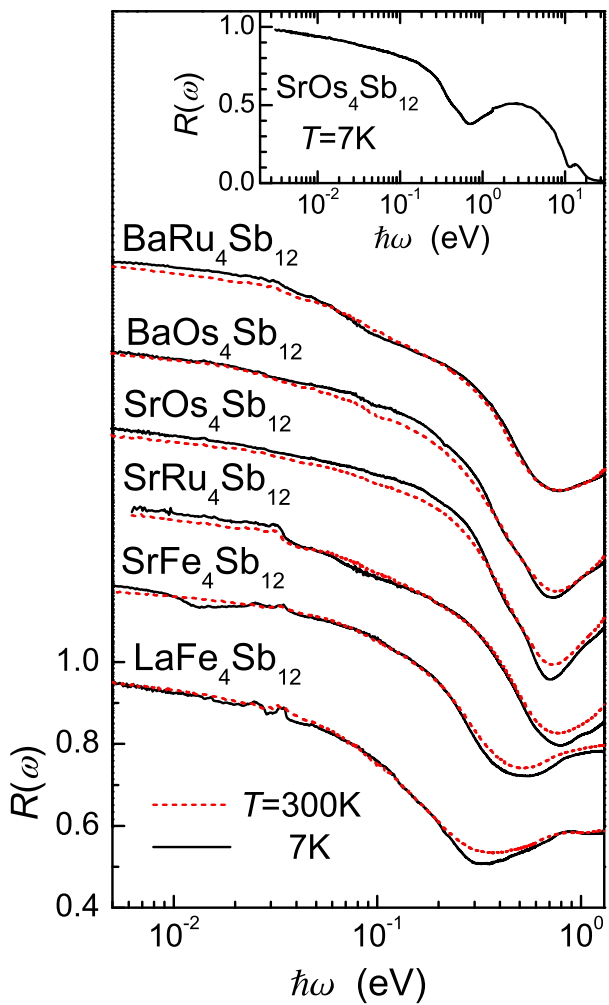}
\end{center}
\caption{
(Color online)
Reflectivity spectra [$R(\omega)$] of BaRu$_{4}$Sb$_{12}$, BaOs$_{4}$Sb$_{12}$, SrOs$_{4}$Sb$_{12}$, SrRu$_{4}$Sb$_{12}$, SrFe$_{4}$Sb$_{12}$ and LaFe$_{4}$Sb$_{12}$ at 7 (solid lines) and 300~K (dashed lines).
Successive curves are offset by 0.2 for clarity.
The whole reflectivity spectrum of SrOs$_4$Sb$_{12}$ in the energy range of 2~meV~--~30~eV is plotted in the inset.
}
\label{reflectivity}
\end{figure}
\begin{figure}[t]
\begin{center}
\includegraphics[width=0.45\textwidth]{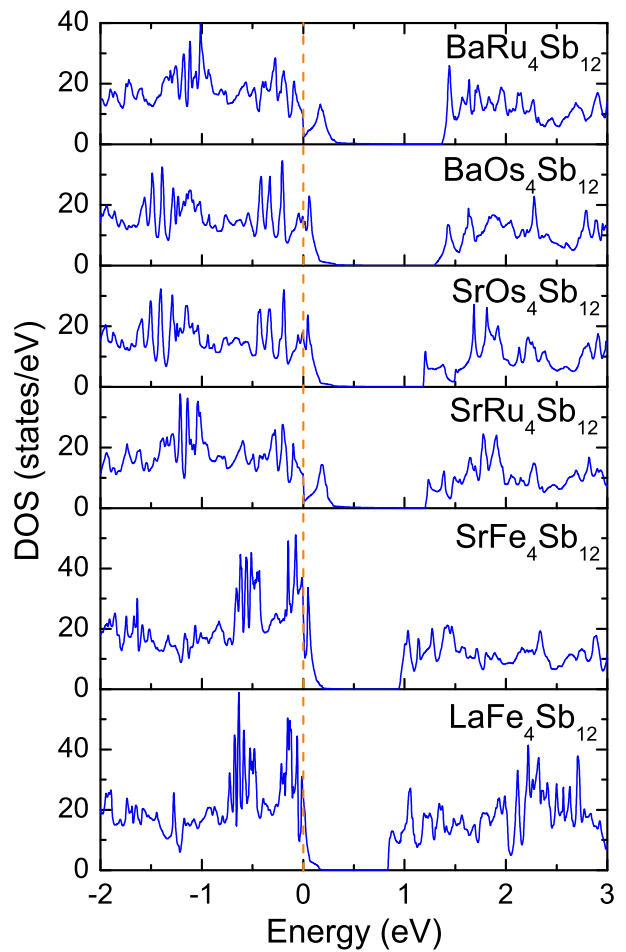}
\end{center}
\caption{
(Color online)
The calculated density of states (DOS) of LaFe$_{4}$Sb$_{12}$, SrFe$_{4}$Sb$_{12}$, SrRu$_{4}$Sb$_{12}$, SrOs$_{4}$Sb$_{12}$, BaOs$_{4}$Sb$_{12}$ and BaRu$_{4}$Sb$_{12}$.
The Fermi energy is denoted by a dashed line.
}
\label{DOS}
\end{figure}
The obtained $R(\omega)$ spectra at 7 and 300~K are shown in Figure~\ref{reflectivity}.
The plasma edge ($\hbar\omega_{p}$) that is the reflectivity minimum in this energy range is located at 0.31~eV in LaFe$_{4}$Sb$_{12}$ and at 0.51~eV in SrFe$_{4}$Sb$_{12}$.
Since $(\hbar\omega_{p})^{2}$ is proportional to the carrier density in the Drude formula, the carrier density of SrFe$_{4}$Sb$_{12}$ is higher than that of LaFe$_{4}$Sb$_{12}$.
The origin of the difference in carrier density originates from the different valence numbers of the guest atoms between these materials, specifically, Sr$^{2+}$ in SrFe$_{4}$Sb$_{12}$ and La$^{3+}$ in LaFe$_{4}$Sb$_{12}$.
The increase in carrier density with the increasing valence number of the guest atoms indicates that positive carriers (holes) are dominant.
This is consistent with the band structure calculations as shown in Figure~\ref{DOS} and also with the positive value in the thermopower and Hall coefficient data in the normal state.~\cite{Taka06}
The band structure calculations indicate that the overall band structure does not change in these materials, with the $E_{\rm F}$ of LaFe$_{4}$Sb$_{12}$ only shifting by about 30~meV from that of SrFe$_{4}$Sb$_{12}$.
It was confirmed that in Na$^{+}$Fe$_{4}$Sb$_{12}$ and K$^{+}$Fe$_{4}$Sb$_{12}$, $\hbar\omega_{p}$ is located at 0.6~eV, which is a higher energy than that of SrFe$_{4}$Sb$_{12}$.~\cite{Joerg07}
Therefore the carrier density of Na- and K-compounds is higher than that of Sr$^{2+}$-compounds.
This indicates that the electrons released from the guest atoms control the $E_{\rm F}$ in the rigid-band-like electronic structure of the Fe$_{4}$Sb$_{12}$ cage.
It should be noted that in comparing with the band structure among $A$Fe$_{4}$Sb$_{12}$ ($A$~=~La, Sr, Na) materials, the hybridization band between the Fe~$3d$ and Sb~$5p$ orbitals commonly exists neat the $E_{\rm F}$, and that the $E_{\rm F}$ is determined only by the valence number of the guest atom.
This indicates that the $R(\omega)$ minimum is not only the carrier based plasma edge, but also includes interband transitions near the $E_{\rm F}$, as discussed later.
This fine structure near the $E_{\rm F}$ is smeared out by thermal broadening at 300~K.
In this case, the minimum $R(\omega)$ value can be regarded as the carrier plasma edge at 300~K.
The spectral weight ($\frac{4m_0}{h^2e^2} \int^{\hbar\omega_p}_{0}\sigma(\hbar\omega)d\hbar\omega$, where $m_0$ the rest mass of an electron and $e$ is the electron charge.) below the plasma edge ($\hbar\omega_p$~=~0.51~eV) of SrFe$_{4}$Sb$_{12}$ is evaluated to be 2.7$\times$10$^{21}$~cm$^{-3}$, which is consistent with the carrier density of CaFe$_{4}$Sb$_{12}$ evaluated by the Hall coefficient measurement.~\cite{Taka06} 

\begin{figure}[t]
\begin{center}
\includegraphics[width=0.45\textwidth]{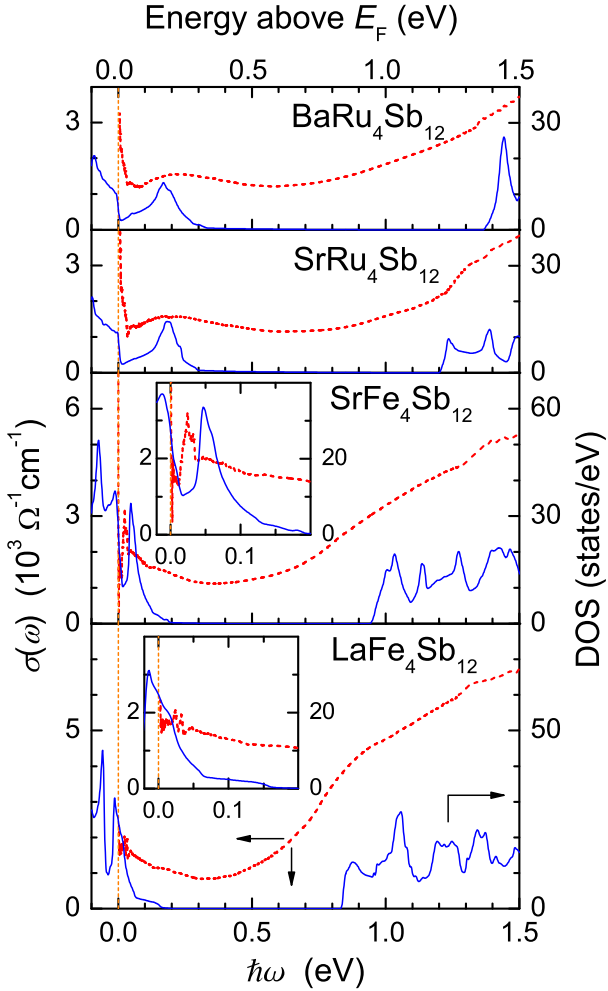}
\end{center}
\caption{
(Color online)
Optical conductivity [$\sigma(\omega)$] spectra (dashed lines) at $T$~=~7~K and the densities of states (DOS, solid lines) of BaRu$_{4}$Sb$_{12}$, SrRu$_{4}$Sb$_{12}$, SrFe$_{4}$Sb$_{12}$ and LaFe$_{4}$Sb$_{12}$.
The $\sigma(\omega)$ spectra roughly correspond to the unoccupied states in the DOS.
The insets in LaFe$_{4}$Sb$_{12}$ and SrFe$_{4}$Sb$_{12}$ are a larger scale of each figure below 0.2~eV.
}
\label{OCvsDOS}
\end{figure}

The next step in this comparison is to make clear which electronic structure an $\sigma(\omega)$ spectrum corresponds to.
The $\sigma(\omega)$ spectra at $T$~=~7~K and the densities of states (DOS) above the $E_{\rm F}$'s of LaFe$_{4}$Sb$_{12}$, SrFe$_{4}$Sb$_{12}$, SrRu$_{4}$Sb$_{12}$ and BaRu$_{4}$Sb$_{12}$ are then plotted, as shown in Figure~\ref{OCvsDOS}.
For LaFe$_{4}$Sb$_{12}$, the $\sigma(\omega)$ spectrum monotonically decreases with increasing photon energy for energies less than 0.4~eV with the exceptions of the TO-phonon structure at around 0.02~eV.
Above 0.4~eV, the $\sigma(\omega)$ spectrum is dominated by the shoulder structure at around 0.8~eV.
The DOS above the $E_{\rm F}$ also monotonically decreases with increasing energy below 0.15~eV, disappears between 0.2 -- 0.8~eV, and appears again above 0.8~eV.
In this case, the $\sigma(\omega)$ spectrum can be attributed to the DOS above the $E_{\rm F}$ in which the spectrum originates from the transition from the DOS at the $E_{\rm F}$ to the unoccupied states.

In the case of SrFe$_{4}$Sb$_{12}$, a shoulder structure (unclear, but it is not flat) in the $\sigma(\omega)$ spectrum appears at around $\hbar\omega=$0.9~eV that corresponds to that at 0.8~eV in LaFe$_{4}$Sb$_{12}$.
In addition, an additional peak in the $\sigma(\omega)$ appears at 25~meV.
Roughly speaking, the additional peak can also be explained by unoccupied states in the DOS, as shown in the inset of Figure~\ref{OCvsDOS}.
However, the peak in the $\sigma(\omega)$ spectrum is located at 25~meV as compared with the location of the DOS peak at 50~meV.~\cite{sk06-2}
The origin of this difference is discussed later.
Due to the rigid-band-like shift of the $E_{\rm F}$, the peak structure that is located below the $E_{\rm F}$ in LaFe$_{4}$Sb$_{12}$ is pushed up above the $E_{\rm F}$ in SrFe$_{4}$Sb$_{12}$.
Coincidently, the ``V''-shaped DOS peak located 40~meV below the $E_{\rm F}$ in LaFe$_{4}$Sb$_{12}$ is moved up to the $E_{\rm F}$.
$\sigma(\omega)$ spectra with a narrow pseudogap structure, which commonly appears in $A^{2+}$Fe$_{4}$Sb$_{12}$ materials, reflect the ``V''-shape DOS of the $E_{\rm F}$.~\cite{Joerg06, sk06-1}

In the case of SrRu$_{4}$Sb$_{12}$ and BaRu$_{4}$Sb$_{12}$, a broad peak in the $\sigma(\omega)$ spectra appears at 0.2~eV that is consistent with the peak in unoccupied states in the DOS as shown in Figure~\ref{OCvsDOS}.
In addition, shoulder structures in the $\sigma(\omega)$ spectra appear at 1.25 and 1.35~eV in SrRu$_{4}$Sb$_{12}$ and BaRu$_{4}$Sb$_{12}$, respectively, that corresponds to additional states of the DOS.
Based on this analysis, it was concluded that the $\sigma(\omega)$ spectra of LaFe$_{4}$Sb$_{12}$, SrFe$_{4}$Sb$_{12}$, SrRu$_{4}$Sb$_{12}$ and BaRu$_{4}$Sb$_{12}$ shown in Figure~\ref{OCvsDOS} correspond to unoccupied states in the DOS.

In the Os-compounds, the $\sigma(\omega)$ spectra can not be explained by these unoccupied states.
According to Figure~\ref{DOS}, DOS fine structures exist not only above but also below the $E_{\rm F}$ for these materials.
It is then possible for the occupied states to also determine the $\sigma(\omega)$ spectra.
An additional comparison of the calculated $\sigma(\omega)$ spectra from the band structure calculations to the experimental $\sigma(\omega)$ spectra is then required.

\begin{figure}[t]
\begin{center}
\includegraphics[width=0.45\textwidth]{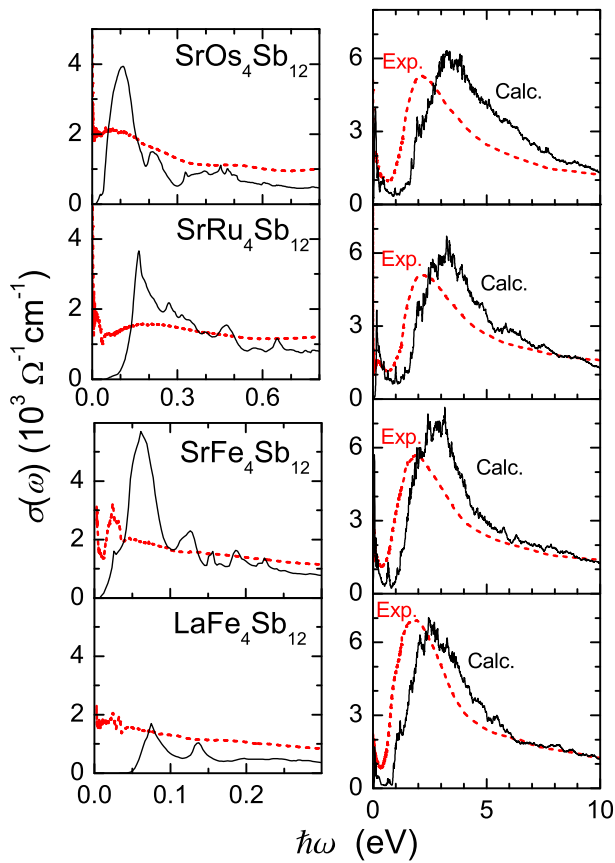}
\end{center}
\caption{
(Color online)
Experimental optical conductivity [$\sigma(\omega)$] spectra at $T$~=~7~K (dashed lines) compared with the corresponding calculated ones (solid lines) of Sr$M_{4}$Sb$_{12}$ ($M$~=~Fe, Ru, Os) and LaFe$_{4}$Sb$_{12}$.
The left figures are expansions of the low energy region.
}
\label{CalcOC}
\end{figure}
To clarify the effect of the different atoms in Fe, Ru and Os to the electronic structure in detail, the experimental $\sigma(\omega)$ spectra are compared with the theoretical ones obtained from the band structure calculation.
The $\sigma(\omega)$ spectra due to interband transitions are derived from the function as follow in which the direct transition is assumed;~\cite{Ant04}
\[
\sigma(\omega) = \frac{\pi e^2}{m_0^2 \omega} \sum_{\vec{k}} \sum_{n n'} \frac{|\langle n' \vec{k}|\vec{e} \cdot \vec{p}|n \vec{k}\rangle |^{2}}{\omega - \omega_{n n'}(\vec{k})+i\Gamma} \times \frac{f(\epsilon_{n\vec{k}})-f(\epsilon_{n'\vec{k}})}{\omega_{n n'}(\vec{k})}
\]
Here, the $|n' \vec{k}\rangle$ and $|n \vec{k}\rangle$ states denote the occupied and unoccupied states, respectively, $f(\epsilon_{n\vec{k}})$ is the Fermi-Dirac function, $\hbar\omega_{n n'}=\epsilon_{n\vec{k}}-\epsilon_{n'\vec{k}}$ is the energy difference of the unoccupied and occupied states and $\Gamma$ is the lifetime parameter.
In the calculation, $\Gamma$~=~1~meV was assumed.

The calculated $\sigma(\omega)$ spectra of Sr$M_{4}$Sb$_{12}$ ($M$~=~Fe, Ru, Os) and LaFe$_{4}$Sb$_{12}$ and the corresponding experimental spectra are shown in Figure~\ref{CalcOC}.
The experimental spectra over the wide energy range in the right figures of Figure~\ref{CalcOC} has a tail towards the higher energy side, which was reproduced by the calculation with reasonable accuracy.
However, the large peak experimentally observed at around $\hbar\omega$~=~2~eV that corresponds to the interband transition from the valence to conduction bands is located at about 3~eV in the calculated spectra.
The experimental peak slightly shifts to the higher energy side changing from $M$~=~Fe to Os, and the corresponding peak in LaFe$_{4}$Sb$_{12}$ is located at a lower energy than that of SrFe$_{4}$Sb$_{12}$.
The shifts in the peak energies are consistent with the calculations.
By comparison, the spectral features in the lower energy region shown in the left figures in Figure~\ref{CalcOC} drastically change with changing $M$.
In the case of SrFe$_{4}$Sb$_{12}$, as reported in previous papers,~\cite{Joerg06,sk06-2} a pseudogap appears below 14~meV, and a peak is present at around 25~meV.
In comparing the calculated spectra with LaFe$_{4}$Sb$_{12}$, the $\sigma(\omega)$ spectrum of SrFe$_{4}$Sb$_{12}$ has a peak at 60~meV.
This peak was assumed to correspond to the experimental peak at 25~meV despite the large energy difference.
The origin of this inconsistency is discussed later.

In the case of SrRu$_{4}$Sb$_{12}$, a broad peak was experimentally observed at 0.2~eV.
There is a corresponding peak at the same energy in the calculated spectrum.
In the case of SrOs$_{4}$Sb$_{12}$, two peaks at 0.1 and 0.5~eV and a shoulder at 0.22~eV were experimentally observed.
All of these structures were reproduced by the calculation.
BaOs$_{4}$Sb$_{12}$ exhibited the same result as SrOs$_{4}$Sb$_{12}$ (results not shown).
Based on this, all of the experimental $\sigma(\omega)$ spectra of the $AM_{4}$Sb$_{12}$ materials reported here can be qualitatively reproduced by the band calculation.
In comparison with the other experiments, the order of the peak energies for $M$ is the same as for the other physical properties of these materials.
For instance, the electronic specific heat coefficient $\gamma$ is 100, 10 and 45 mJ/mol~K$^2$ in $M$~=~Fe, Ru, Os, respectively.
This indicates that the low energy peak in the $\sigma(\omega)$ spectra as well as the shape of the DOS near the $E_{\rm F}$ strongly affects the thermodynamical physical properties in $AM_{4}$Sb$_{12}$ materials.


\begin{figure}[t]
\begin{center}
\includegraphics[width=0.4\textwidth]{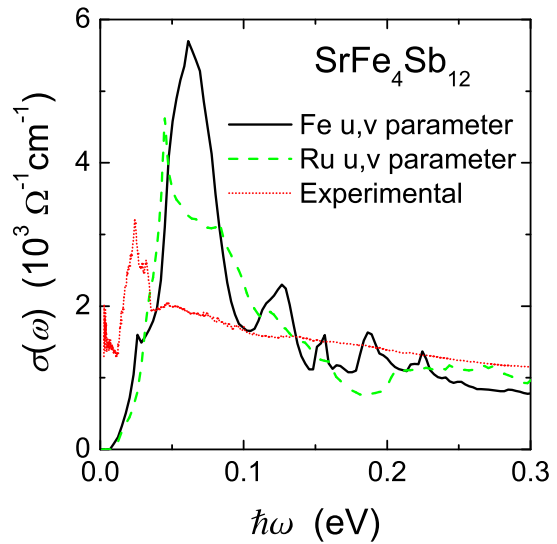}
\end{center}
\caption{
(Color online)
Calculated optical conductivity [$\sigma(\omega)$] spectra of SrFe$_{4}$Sb$_{12}$ using the positional parameters of SrFe$_{4}$Sb$_{12}$ (solid line) and of SrRu$_{4}$Sb$_{12}$ (dashed line) compared with the experimental spectrum (dotted line).
}
\label{ParamDep}
\end{figure}
The possibilities of the origin of the difference of the spectral shapes among $M$~=~Fe, Ru, Os in the infrared region include the different positional parameters of Sb ions and the differences in the wave function of the $d$ electrons due to the different principal quantum numbers.
In order to ascertain the effects of the different positional parameters, the $\sigma(\omega)$ spectrum of SrFe$_{4}$Sb$_{12}$ with the positional parameter of the Ru-counterpart was calculated, as shown in Figure~\ref{ParamDep}.
The characteristic peak at 60~meV is again present even using the Ru positional parameters, although the intensity is lower.
This result indicates that the positional parameters of Sb ions do not affect the electronic structure near the $E_{\rm F}$, but the differences in the wave functions of the $d$ states in $M$~=~Fe, Ru, Os are dominant.
Therefore, the differences in the physical properties among these materials also originate from the different wave functions of the $d$ states.
Particularly in the case of SrFe$_{4}$Sb$_{12}$, the electronic structure near the $E_{\rm F}$ is largely determined by the Fe~$3d$~--~Sb~$5p$ hybridization state.
The pseudogap structure is concluded to originate from the localization of the Fe $3d$ state.
As noted previously, the experimentally obtained peak energy of 25~meV for SrRu$_{4}$Sb$_{12}$ is lower than the calculated one (60~meV), which is not the case for the other compounds.
This low energy shift in the experimental peak might be due to the renormalization effect arising from the strong electron correlation or due to the self-energy effect observed in the photoemission spectra.~\cite{Kord05}
The existence of heavy quasiparticles originating from the Fe $3d$ state has been previously reported.~\cite{sk06-1}
The origin of the heavy quasiparticles is believed to be the Kondo effect resulting from the hybridization between the localized Fe $3d$ state and conduction band.
However, since SrFe$_{4}$Sb$_{12}$ is located very near the ferromagnetic ordering state, the spin fluctuation effect must also be considered.
Heavy quasiparticles due to the spin fluctuation are also predicted by the SCR theory.~\cite{Hasegawa79, Moriya91}
Both the peak shift and the presence of heavy quasiparticles suggest the Fe~$3d$ state in SrFe$_{4}$Sb$_{12}$ has stronger electron correlation than the Ru- and Os-compounds.

\section{Conclusion}

In summary, to investigate the electronic structure of the $M_{4}$Sb$_{12}$ cage in $AM_{4}$Sb$_{12}$ ($A$~=~Sr, Ba, La, $M$~=~Fe, Ru, Os) materials, we measured the optical conductivity [$\sigma(\omega)$] spectra and compared with band structure calculations as well as calculated $\sigma(\omega)$ spectra.
Both the energy shift of the plasma edge due to the different valence number of guest atoms $A$ in the Fe$_{4}$Sb$_{12}$ cage and the corresponding band calculation suggest the hole-type carriers.
The experimental $\sigma(\omega)$ spectra were reasonably well reproduced by the band calculations using experimental lattice constants and positional parameters.
The $\sigma(\omega)$ spectra of Sr$M_{4}$Sb$_{12}$ ($M$~=~Fe, Ru, Os) in the infrared region drastically change with different $M$.
The origin of this change does not originate from the different positional parameters of Sb ions but from the different wave functions of $d$ states due to the different principal quantum numbers of $M$ ions.

\section*{ACKNOWLEDGMENT}

We would like to thank Prof. K. Takegahara for the fruitful discussion and suggestions.
This work was a joint studies program of the Institute for Molecular Science (2005) and was partially supported by a Grant-in-Aids; the COE Research (Grant No.~13CE2002), the priority area "Skutterudite" (No.~15072205) and Scientific Research (B) (No.~18340110) from MEXT of Japan.


\end{document}